# The Kondo effect of surface excitons

P. D. Altukhov*

A. F. Ioffe Physical-Technical Institute, St. Petersburg, Russia

*Corresponding author:

P. D. Altukhov,

A. F Ioffe Physical-Technical Institute, Politekhnicheskaya street 26, St. Petersburg 194021, Russia;

Tel: +7-812-292-7344;

Fax: +7-812-297-1017;

E-mail: pavel.altukhov@gmail.com

**ABSTRACT**

A recombination radiation line of real excitons in dense two-dimensional electron gas at the [100] silicon surface is observed in luminescence spectra of metal-oxide-semiconductor (MOS) structures. A new effect of anisotropic paramagnetic reduction of the luminescence line indicates a strong influence of the Kondo correlations on electron paramagnetism of the excitons.

## 1. INTRODUCTION

The electron-hole interaction in presence of metallic electrons manifests itself as the Fermi-edge singularity in recombination radiation spectra of metals [1] and two-dimensional (2D) electrons in semiconductor quantum wells [2]. The Fermi-edge singularity is a many-body phenomenon and can be explained in terms of multiple electron-hole scattering as a result of an increase of the electron density at the Fermi level near a hole [1]. A similar singularity is observed in the recombination radiation line (S-line) of 2D electrons and 2D nonequilibrium holes, bound to the 2D electron layer by the electron polarization attraction at the [100] silicon surface in MOS structures [3,4]. Moreover, real excitons are observed in the S-line spectrum at high densities of 2D electrons [4]. Changing of the 2D electron density can change the distance of the excitons from the surface and interaction between the excitons and 2D electrons. Presented here magnetooptical experiments show, how the surface exciton transforms itself from a real single-particle state to a many-particle virtual bound state and how the Kondo correlations [5] influence electron paramagnetism of the surface excitons.

## 2. RESULTS AND DICUSSION

The spectral position of the high energy edge of the S-line is very close to the spectral position of the bound exciton line (BE) and varies weakly with increasing 2D electron density $n_S$ (Figs 1 and 2). This implies, that the energy level of bulk excitons in silicon is close to the Fermi level of 2D electrons. The S-line spectrum reproduces the energy spectrum of 2D electrons at $n_S \approx (0.5 \div 1.5) \times 10^{12} cm^{-2}$. In this region surface electron-hole pairs exist as a 2D two-layer electron-hole plasma, and an additional 2D hole layer is formed [3,4]. At low pair density holes in the additional layer are non-degenerate and occupy the lowest quantum level with the spin momentum ±3/2 in the hole quantum well. At $n_S < 0.5 \times 10^{12} cm^{-2}$ surface pairs exist as excitons due to weak *k*-space occupation and weak screening [3,4].

At $n_S > 1.5 \times 10^{12} cm^{-2}$ the Fermi-edge singularity, increasing with increasing $n_S$, arises at the high energy edge of the S-line (Figs 1 and 2) due to an increase of the radiative transition probability for the electron levels close to the Fermi level. The Fermi-edge singularity at the beginning of its formation can be explained in terms of multiple electron-hole scattering [1] as a result of formation of a many-particle virtual bound state. The Fermi-edge singularity (excitonic resonance) becomes very strong and narrow before the second electron subband occupation, which occurs at $n_S > 6.4 \times 10^{12} cm^{-2}$ [6]. At the same time a new narrow line (S'-line) arises at the high energy edge of the S-line at low temperatures [3,4] (Figs 1 and 2). The observed behavior of the luminescence spectra has to be attributed to formation of real excitons in the system of surface pairs at $n_S \approx (3 \div 6) \times 10^{12} cm^{-2}$. In this region the excitonic resonance in the S-line spectrum corresponds to surface excitons, whose electrons originate from the lowest electron subband. Such an S-exciton, strongly interacting with 2D electrons, can be considered as a real bound state. The S'-line corresponds to surface excitons, whose electrons originate, presumably, from upper subbands. These excitons, arranged away from the 2D electron layer, have the lowest energy (Fig. 2). The maximum energy difference between the S-exciton and the S'-exciton is about 0.3 *meV* [3].

The formation of real excitons in dense 2D electron gas can be explained as a result of a self-organization of surface pairs, possessing an opportunity to choose their distance from the surface, which minimize the pair energy. Combination of surface and bulk states causes the existence of two surface excitonic states. The states of heavy and light holes in the S'-exciton are not split practically due to a large distance of the S'-exciton from the surface. These states in the S-exciton are split by a surface potential. With decreasing $n_S$ the distance of the S-exciton from the surface decreases, resulting in a broadening of the excitonic resonance and in an increase of the low energy tail of the S-line spectrum (Fig. 1). The low energy tail of the S-line spectrum is defined by recombination of a hole and 2D electrons and is caused by a penetration of the 2D electron wave functions into the S-exciton. This penetration becomes strong at low $n_S$, and reaches the maximum at the Fermi level, giving a contribution to the excitonic resonance. An essential contribution to the excitonic resonance is given by the radiative recombination of a hole and "its own electron".

A new effect of anisotropic paramagnetic reduction of the excitonic resonance in a tilted magnetic field [4] (Figs 1,2 and 3) proves the existence of real excitons in dense 2D electron gas. This effect is caused by a decrease of the average radiative transition probability in the magnetic field due to simultaneous occupation of the lowest electron and hole spin states with the spin momentum −1/2 and −3/2 and the radiative transition probability equal to zero. 2D electrons are not orientated in the magnetic field due to a large Fermi energy $E_F$. So, the effect of paramagnetic reduction is a property of excitons, because electrons in excitons can be orientated in a magnetic field independently from 2D electrons. The paramagnetic reduction of the excitonic resonance depends only on the transverse component of magnetic field $B_{\perp}$. The paramagnetic reduction of the excitonic resonance is absent in the longitudinal magnetic field $B_{||}$. The strong anisotropy of the reduction is induced by a strong anisotropy of the 2D hole g-factor, resulting from a very small value of the 2D hole g-factor in the longitudinal magnetic field [4].

The problem of electron paramagnetism of excitons in dense 2D electron gas is very similar to the problem of impurity paramagnetism in metals [5]. Observation of the Curie paramagnetism of electrons in the excitons is possible when the Kondo temperature of the excitons $kT_K$ is lower than the electronic temperature $kT$, and the Kondo temperature is equal to the Kondo electron spin-correlation frequency of the exciton and 2D electrons. High spin-correlation frequency suppresses the electron paramagnetism of the excitons. The Kondo electron spin singlet of the exciton and 2D electrons is formed at $kT < kT_K << \Delta$ [5], where $\Delta$ is the spectral width of the excitonic resonance, defined by the decay time of the exciton. At $kT_K \approx \Delta$ the excitonic resonance corresponds to the virtual bound state. The magnetic field dependence of the paramagnetic reduction of the excitonic resonance shows the Curie behavior of the electron paramagnetism of the S-excitons at $n_s > 4\times10^{12} cm^{-2}$ (Fig. 3). In this region the amplitude of the reduction depends on the temperature, decreases with decreasing $n_s$, and is described by equation (1) of Ref. [4]. Comparison of the theory [4] and the experiment gives the values of the hole g-factor $g_{h\perp} = 0.5$ and $g_{h||} = 0$. At high transverse magnetic fields recombination radiation of a hole and "its own electron" is eliminated from the excitonic resonance and the S-line spectrum is defined by the radiative recombination of holes and 2D electrons. The hole states in the S'-excitons are not split, the hole g-factor is isotropic, and the S'-line intensity reveals the isotropic paramagnetic reduction (Figs 1,2 and 3).

The Kondo temperature of the excitons increases with decreasing $n_S$, and at $n_S < 4\times10^{12}\,cm^{-2}$ the magnetic field dependence of the excitonic resonance intensity shows a strong deviation of electron paramagnetism of the excitons from the Curie behavior, corresponding to the condition $T_K > T$ (Fig 3, curves 1 and 2). At $T_K >> T$ and $3g_{h\perp}\mu_0 B >> kT$, when holes occupy the lowest spin state, the reduction of the excitonic resonance intensity $I$ can be approximately described by the formula

$$\frac{I}{I_0} \approx 1 + \frac{W_+ - W_-}{W_+ + W_-} 2\{-(g\mu_0 B_\perp / 4kT_K)[1 + (g\mu_0 B_\perp / 2kT_K)^2]^{-1/2}\}. \qquad (1)$$

Here $I_0$ is the excitonic resonance intensity at $B = 0$, $\mu_0$ is the Bohr magneton, $g = 2$ is the electron g-factor. $W_+$ and $W_-$, defined in Ref. 4, are the radiative transition probability for the upper and lower electron spin state and the lower hole spin state. The contents in the brackets represent the average electron spin for the Kondo singlet state [5]. The observed transformation of the excitonic resonance reduction with decreasing $n_S$ shows a picture of transformation of the real exciton into the virtual bound state. A strong influence of the Kondo correlations on hole paramagnetism of excitons and multi-exciton complexes, bound to the [100] 2D hole layer, was observed in Ref. 7. Note, that a weak Zeeman splitting of the excitonic resonance in the region of existence of the Kondo spin singlet is not observed due to high spectral width of the S-line. This defines a difference between the excitonic resonance and the well-known Kondo resonance in conductivity of a semiconductor quantum dot. Zeeman splittings of the S-exciton and the S'-exciton in the region $T_K < T$ are detected as simultaneous shifts of corresponding lines in the magnetic field (Fig. 1) without mixing of these lines.

## 3. CONCLUSION

Excitons, bound to a two-dimensional electron layer at the silicon surface, represent a unique system for investigations of Kondo effects. Variation of the density of two-dimensional electrons gives a change of the Kondo temperature in a wide range due to variation of the distance of the excitons from the surface.

# FIGURE CAPTIONS

**Figure 1.** Recombination radiation spectra of 2D electrons and nonequilibrium holes in [100] silicon MOS sructures at $T = 1.9\ K$ and excitation level $10^{-3}\ W\,cm^{-2}$, TO-lines.

$n_s = 2.9\times10^{12}\,cm^{-2}$ for spectra 1,1'; $n_s = 5.1\times10^{12}\,cm^{-2}$ for spectra 2,3,3',4,4'.

1,2: $B = 0$; 3: $B = B_\perp = 2.7\ T$; 1',4: $B = B_\perp = 6.7\ T$; 3': $B = B_{||} = 2.7\ T$; 4': $B = B_{||} = 6.7\ T$.

**Figure 2.** Dependence of the intensity $I$ of the S-line maximum (1,1') and the S'-line intensity (2,2') on $n_S$ at $T = 1.9\ K$.

1: $B = 0$ and $B = B_{||} = 5.8\ T$; 1': $B = B_\perp = 5.8\ T$; 2: $B = 0$; 2': $B = B_\perp = 5.8\ T$ and $B = B_{||} = 5.8\ T$. 3 is the electron density $n_s^1$ in the second subband [6]. 4 is the spectral position of the high energy edge of the S-line $h\nu_S$.

**Figure 3.** Dependence of the intensity $I/I_0$ of the S-line maximum (1–4,1'–4') and the S'-line intensity (5,5') on the magnetic field $B$ at $T = 1.9\ K$.

1–5: $B = B_\perp$ (black circles); 1'–5': $B = B_{||}$ (open circles).

1,1': $n_s = 3.2\times10^{12}\,cm^{-2}$; 2,2': $n_s = 3.8\times10^{12}\,cm^{-2}$; 3,3': $n_s = 4.3\times10^{12}\,cm^{-2}$; 4,4',5,5': $n_s = 5.1\times10^{12}\,cm^{-2}$; $T_K < T$ for curves 3,3',4,4'; $T_K \approx 5.4\ K$ for curves 2,2'; $T_K \approx \Delta$ for curves 1,1'. Dashed lines 3,4 represent the theory of Ref. 4 with $W_- / W_+ = 0.29$ for 3 and $W_- / W_+ = 0.13$ for 4. Dashed line 2 is the theory, represented by formula (1) with $W_- / W_+ = 0.53$. Dashed line 5 is a theory for the S'-line with the hole g-factor $g_1 = 0.8$.

Figure 1

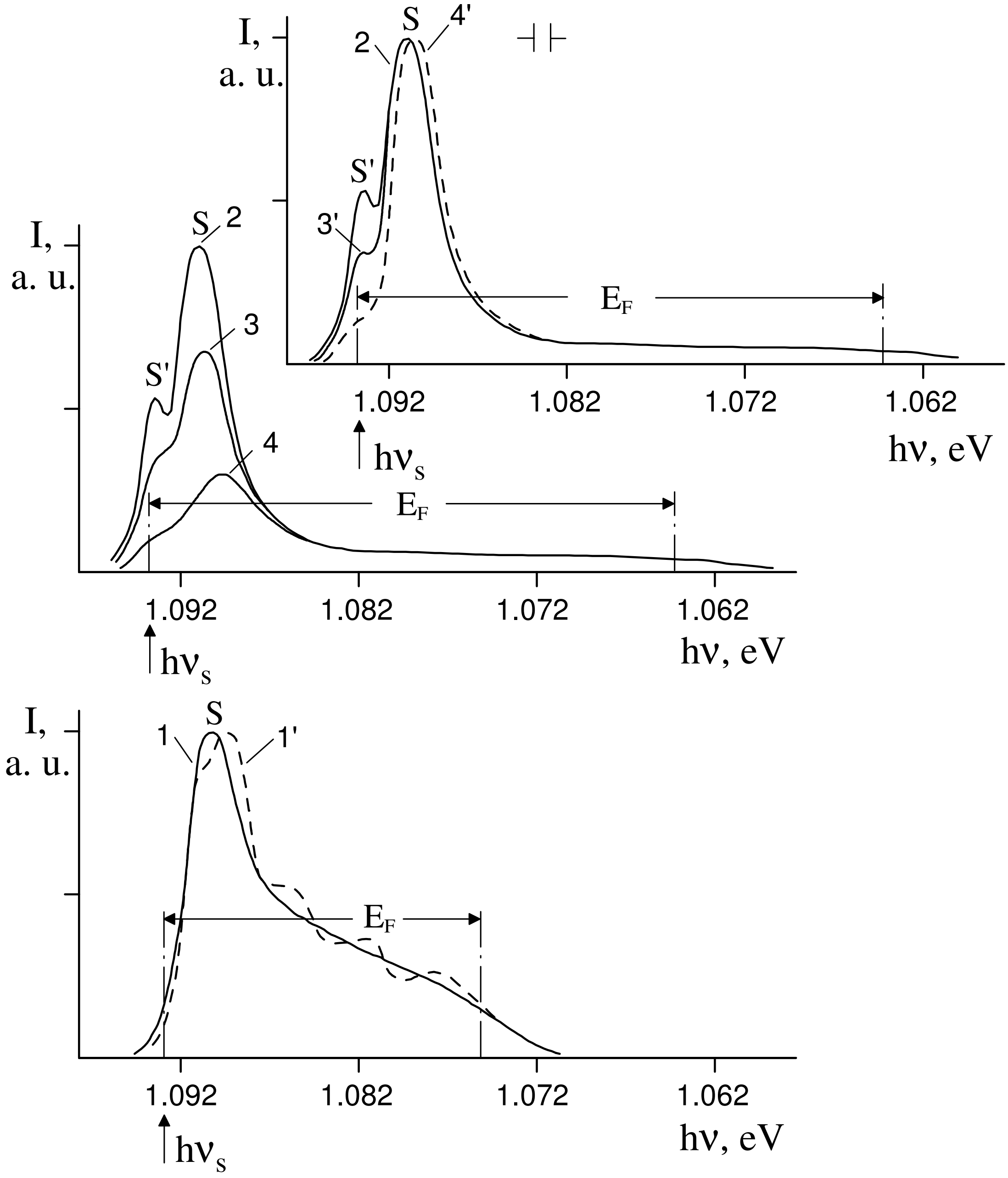

Figure 2

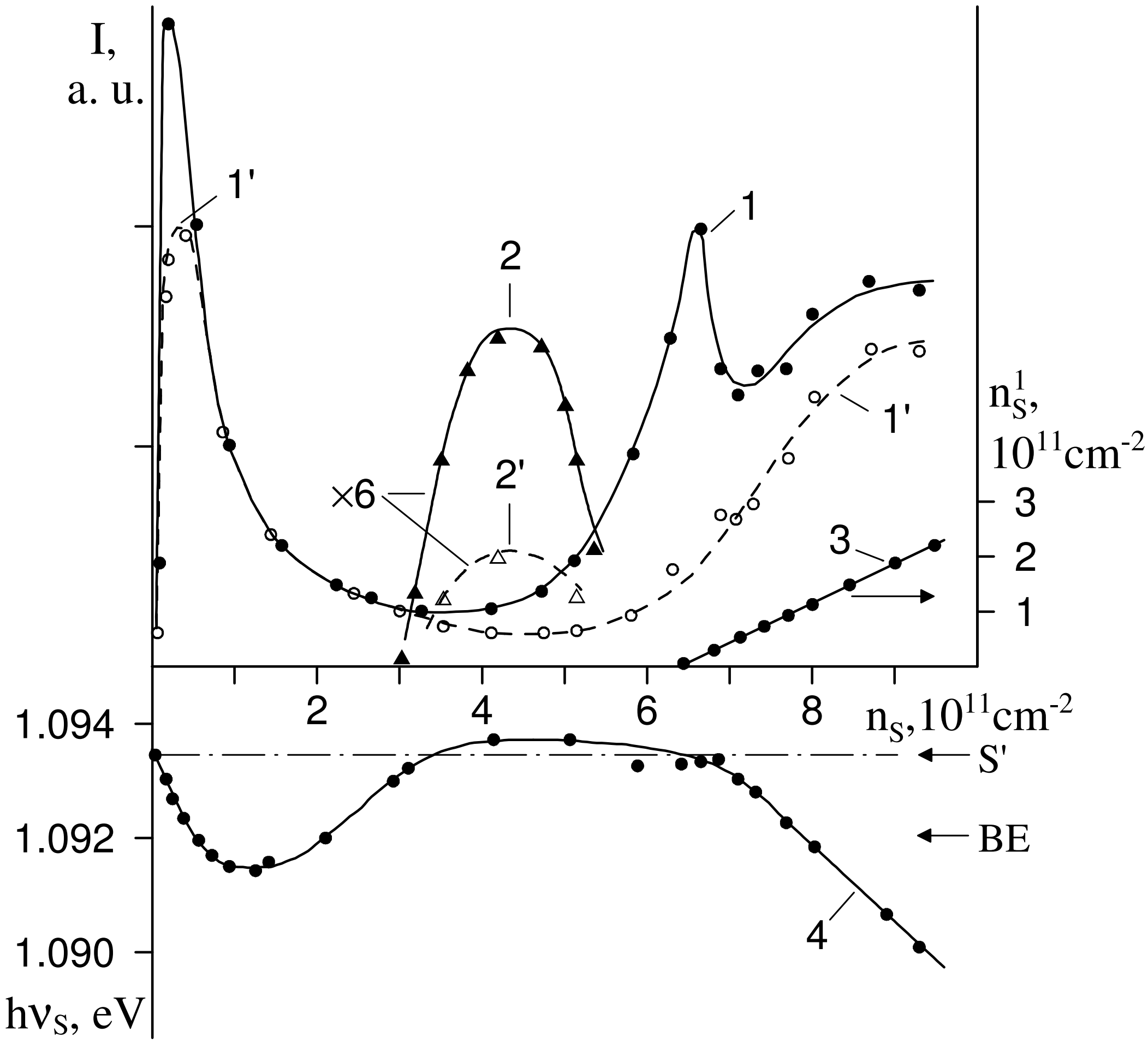

Figure 3

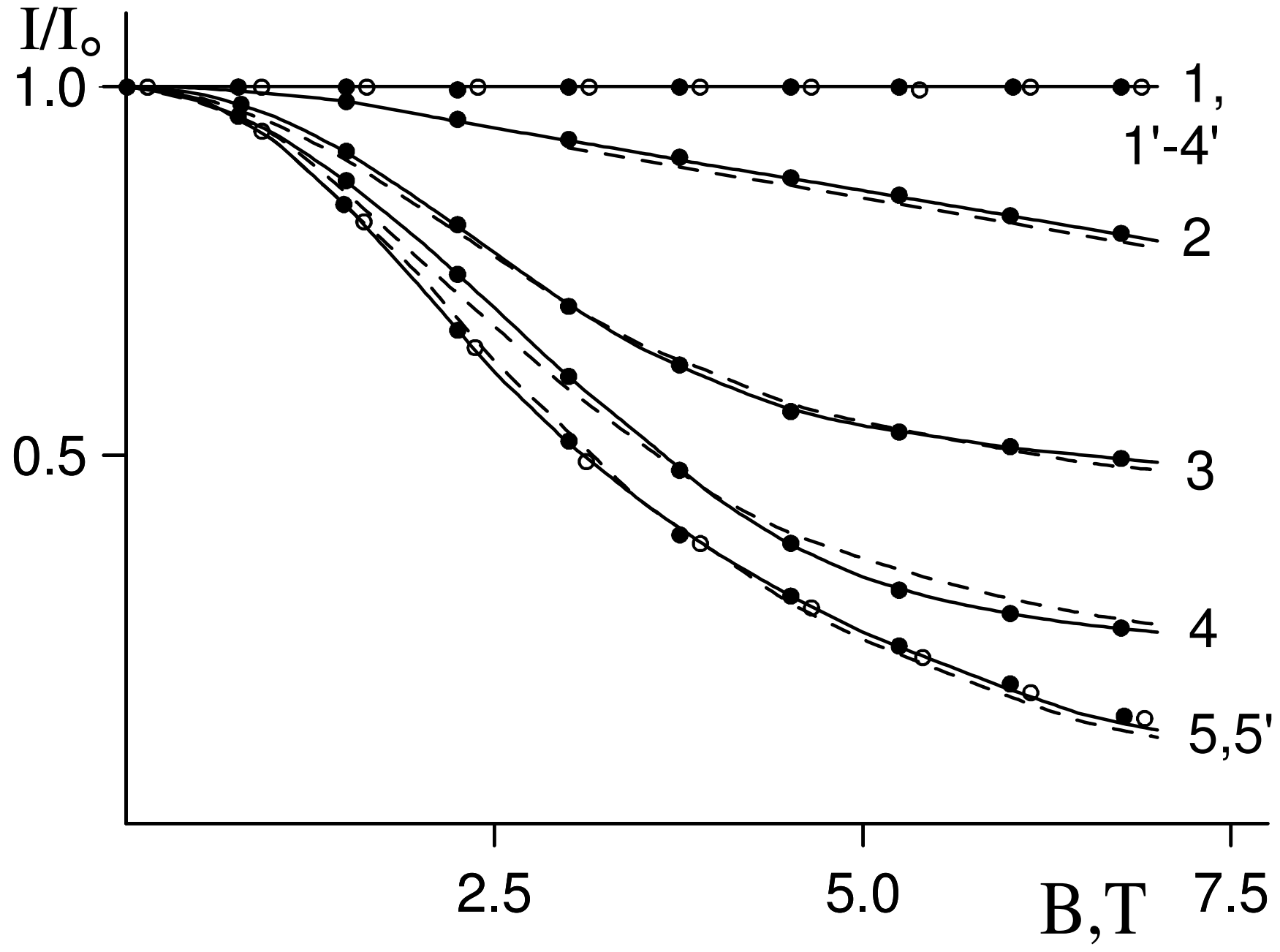